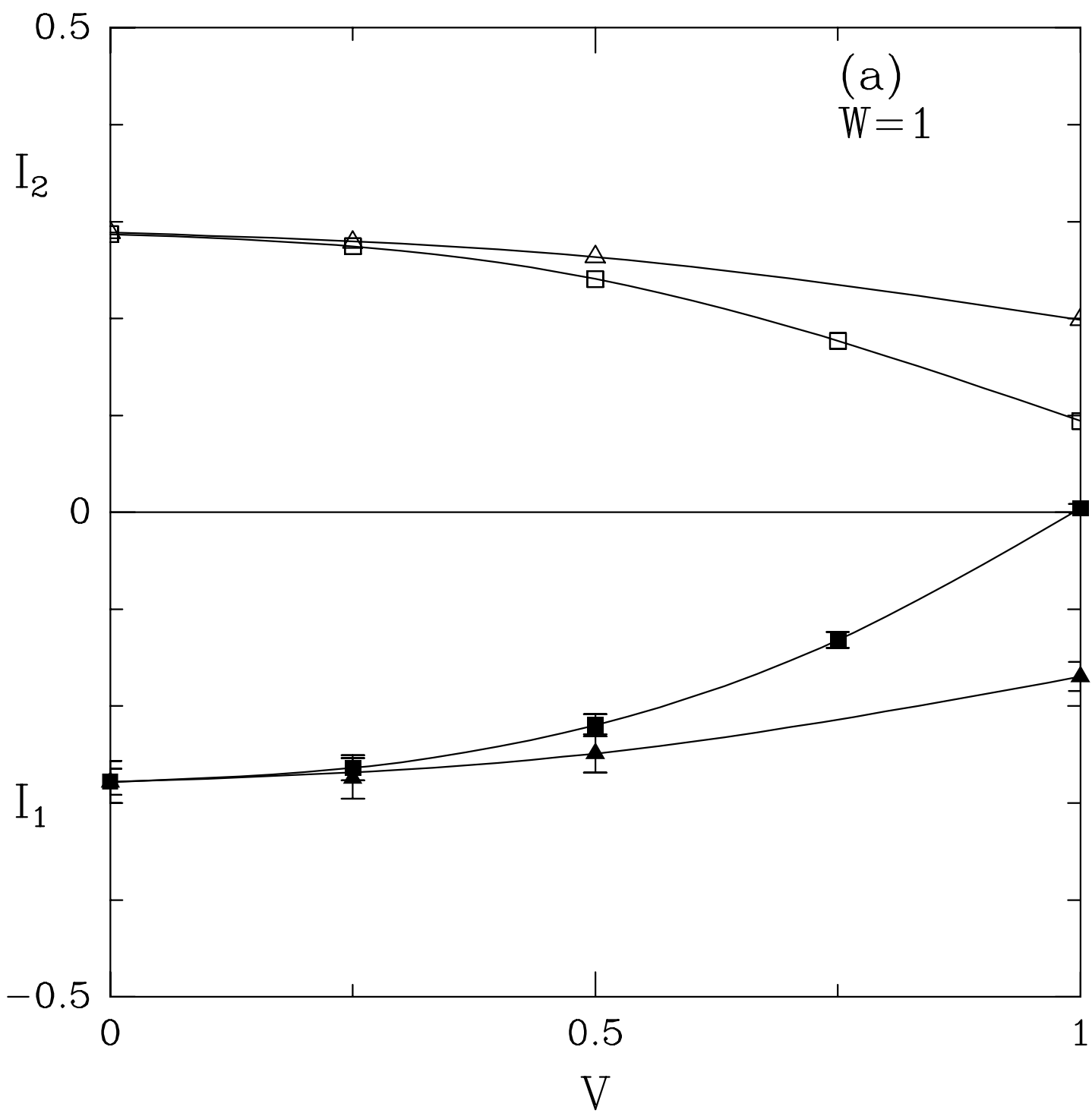

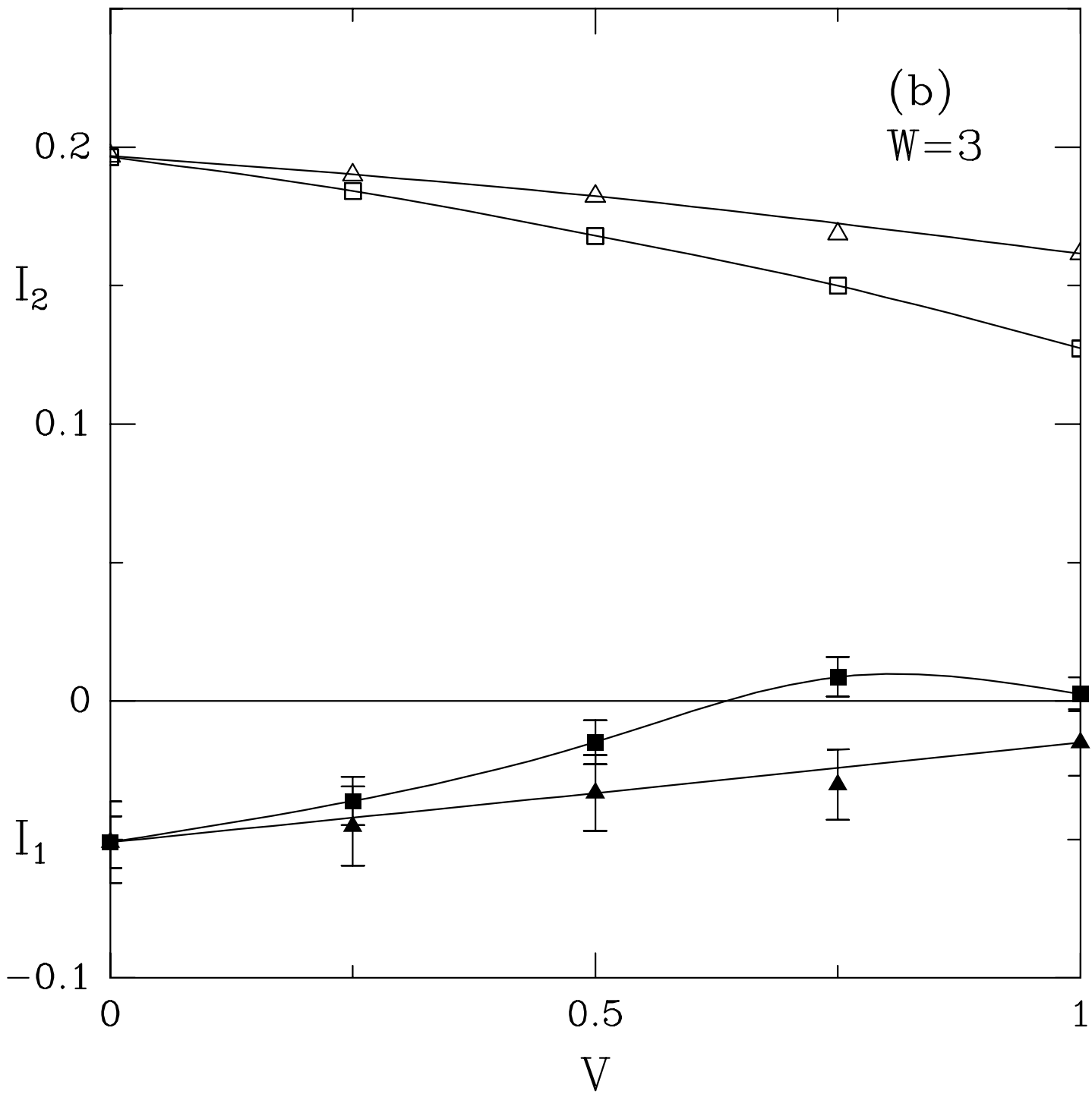

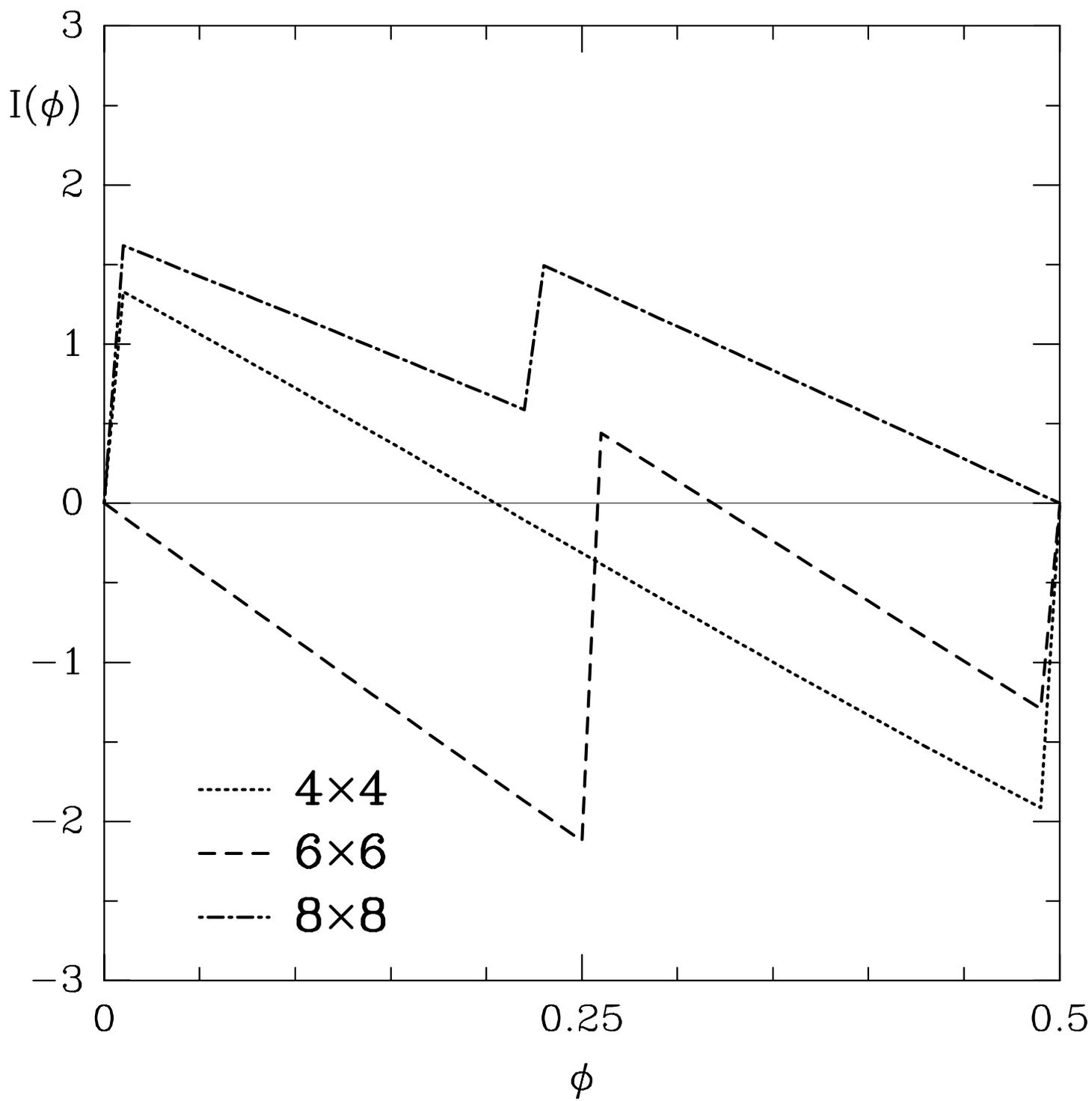

# Persistent currents in multichannel interacting systems

Georges Bouzerar* and Didier Poilblanc**

*Groupe de Physique Théorique, Laboratoire de Physique Quantique,*
*Université Paul Sabatier, 31062 Toulouse, France*

(December 19)

Persistent currents of disordered multichannel mesoscopic rings of spinless interacting fermions threaded by a magnetic flux are calculated using exact diagonalizations and self-consistent Hartree-Fock methods. The validity of the Hartree-Fock approximation is controled by a direct comparison with the exact results on small 4 × 4 clusters. For sufficiently large disorder (diffusive regime), the effect of repulsive interactions on the current distribution is to slightly decrease its width (mean square current) but to *increase* its mean value (mean current). This effect is stronger in the case of a long range repulsion. Our results suggest that the coupling between the chains is essential to understand the large currents observed experimentally.

The first observation of mesoscopic currents in very pure metallic nano-structures was done in pioneering experiments of Levy et al.[1] and Chandrasekhar[2,3]. In the first case the experiment dealt with the average current of a system of $10^7$ disconnected rings in the diffusive regime while in the second a single ring was used. Although the existence of permanent currents in small metallic rings was predicted long ago[4,5] the magnitude of the observed currents is still a real challenge to theorists. There has been a great activity to find a mechanism providing the correct magnitude of the currents. There is a general belief that the interaction has a crucial role in enhancing the current. But, so far, the role of the interaction in disordered systems is still unclear. Treating interaction and disorder on equal footings is a difficult task.

In some previous work[6,7] we have shown by exact diagonalizations (ED) of small clusters that, for strictly 1D systems of spinless fermions, the effect of a *repulsive* interaction is to increase further the localization of the electrons and hence to decrease the value of the current. Using a Hartree-Fock approach, Kato et al.[15] have obtained qualitatively good agreement with our exact calculations. However, Giamarchi et al.[8] have shown that in the case of the 1D Hubbard model (spin is included) the interaction enhances the persistent current. We then expect the increase of the current to be closely related to the *decrease* of the charge fluctuations or, in other words, to the smoothening out of the charge density as this occurs in the 1D Hubbard model whith repulsive interaction. Although this effect does not occur for 1D spinless fermions, it might exist in higher dimensions. We shall therefore consider multi-channel rings relevant to experiments. Unfortunately in the case of multi-channel systems the ED approach[9] is limited to small rings (small number of channels) since the size of the Hilbert space increases exponentially fast with the number of lattice sites. In this paper, we use also an alternative and complementary approach based on a self-consistent Hartree-Fock (HF) treatment of the interaction between particles which enables one to treat much larger systems. Indeed, for increasing number of channels ie increasing dimensionnality quantum fluctuations become less effective and a mean field treatment of the interaction might be appropriate. However, we shall treat the quantum interference due to the disorder somehow exactly. Our method is different from the usual perturbative approach[10] where the corrections to the current due to the interacting term (Hartree and Fock terms) are calculated perturbatively. Recently, Ramin et al.[12] have numerically shown by such a perturbative approach that the first order correction to the second harmonic of the persistent current was in agreement with the analytical treatment[10]; a nearest neighbour interaction tends to decrease the value of the typical current, while a repulsive extended Hubbard interaction enhances the typical current in any dimension. On the contrary, in our self-consistent approach disorder and interaction are treated on equal footings. In another different approach Kopietz[11] by treating the Coulomb repulsion in a semi-classical way has found persistent currents in agreement with those measured by Chandrasekhar et al.[2].

This paper is organized as follow: first, the validity of the mean field treatment is controlled by comparing the HF results to the exact data obtained by Lanczos diagonalizations of small clusters. On the basis of this comparison an accurate estimation of the current in the HF approximation is suggested. Secondly, we use the HF method to handle larger clusters. We show that the interaction modifies the distribution of the currents. In the ballistic regime (small disorder) the interactions have little effects on the currents. However, in the diffusive regime, Coulomb repulsion between fermions can produce a significant increase of the mean-current. We do not observe such a systematic effect in the case of a short range potential.

The Hamiltonian reads:

$$\mathcal{H} = -\sum_{\mathbf{i},\mathbf{j}} t_{\mathbf{ij}} c_{\mathbf{i}}^{\dagger} c_{\mathbf{j}} + \frac{1}{2} \sum_{\mathbf{i},\mathbf{j}} V_{\mathbf{ij}} n_{\mathbf{i}} n_{\mathbf{j}} + \sum_{\mathbf{i}} w_{\mathbf{i}} n_{\mathbf{i}} \qquad (1)$$

where $\mathbf{i},\mathbf{j}$ are vectors of a 2D lattice (x,y) with $0 \leq x \leq L_x - 1$ and $0 \leq y \leq L_y - 1$, $L_y$ being the number of channels, $L_y \leq L_x$. $n_{\mathbf{i}}$ is the local density and $w_{\mathbf{i}}$ are on-site energies chosen randomly between -W/2 and W/2. In the following, we shall consider the cases of (i) a short range repulsion, $V_{\mathbf{ij}} = V \delta(|\mathbf{i} - \mathbf{j}| - 1)$, and (ii) a Coulomb repulsion, $V_{\mathbf{ij}} = V/|\mathbf{i} - \mathbf{j}|$, $\mathbf{i} \neq \mathbf{j}$. The hopping terms $t_{\mathbf{ij}} = \frac{1}{2} \exp(2i\pi\Phi_{\mathbf{ij}})$ are restricted to the bonds connecting nearest neighbor sites. Periodic boudary conditions are assumed in both directions. A flux $\Phi$ (in units of $\Phi_0 = hc/e$) is applied through one hole of the 2D torus leading to a twist in the boundary conditions along x. This is realized e.g. by choosing $\Phi_{\mathbf{ij}} = \frac{\Phi}{L_x} \hat{\mathbf{u}}_{\mathbf{x}} \cdot (\mathbf{i} - \mathbf{j})$ in (1). Hereafter, when not specified, energies are measured in unit of t.

The current density operator is defined by $J_x = -\frac{1}{2\pi} \frac{\partial \mathcal{H}(\Phi)}{\partial \Phi}$ or

$$J_x = \frac{it}{2L_x} \sum_{\mathbf{i}} (\exp(2i\pi\Phi/L_x) c_{\mathbf{i}}^{\dagger} c_{\mathbf{i}+\hat{\mathbf{u}}_{\mathbf{x}}} - h.c.), \qquad (2)$$

and the current in the ring is naturally given by $I(\Phi) = \frac{1}{2\pi} \langle \frac{\partial \mathcal{H}(\Phi)}{\partial \Phi} \rangle$ where $\langle \rangle$ means the expectation value in the ground state. Note that it can also be expressed alternatively as $I(\Phi) = -\frac{1}{2\pi} \frac{\partial E(\Phi)}{\partial \Phi}$ where $E(\Phi)$ stands for the exact GS energy.

Due to the random nature of the disorder, one can characterize the response to the flux by the *statistical distribution* of the currents. Calculations are made for a fixed set of the parameters (V,W,$\phi$) and we average over at least 500 realisations of the disorder (w). The current distribution is investigated by calculating its first and second moments $I_1 = \langle I(\Phi) \rangle_w$ and $I_2 = \sqrt{\langle I^2(\Phi) \rangle_w}$, where $\langle \rangle_w$ means the average over disorder. Note that throughout the paper the moments of the currents are plotted in unit of $I_0$, average over half a period of $I^2(\Phi)$ for $V = 0$ and $W = 0$. The electron density is fixed to $\langle n \rangle = 1/4$ but we expect our results to be generic of the low density limit.

As a first step, exact diagonalisations (ED) of (1) on a small $4 \times 4$ cluster are performed using a Lanczos algorithm[13] to calculate $E(\Phi)$. Here we set $\Phi = 1/4$. Because of the limited length $L_x = 4$ of the ring we restrict ourselves here to a repulsion between only nearest neighbor sites. In fig.1(a) and (b) the first and second moments of the distribution of $I(\Phi)$ are plotted versus V. We observe that the interactions do not enhance the current in this regime of disorder. Note that for such values of W the system is still in the ballistic regime. Indeed, the localization length is $\xi \sim L_y l_e$, where $L_y$ is the number of channels. If one estimates the elastic mean free path as $l_e \sim \frac{100}{(2W/t)^2}$[14] then for $W/t \leq 4$ one gets $\xi > L_x$. A diffusive motion of the electrons is then difficult to realize in such a small ring. On the contrary, for strong impurity scattering (localized regime) an increase of the currents was found[9]. These exact data will become very useful in the following as a necessary reference to control our approximation scheme.

We now turn to the HF approximation of (1). The HF hamiltonian reads

$$\mathcal{H}_{hf} = -\sum_{\mathbf{i},\mathbf{j}} t_{\mathbf{ij}}^{eff} c_{\mathbf{i}}^{\dagger} c_{\mathbf{j}} + \sum_{\mathbf{i}} w_{\mathbf{i}}^{eff} n_{\mathbf{i}}$$
$$- \frac{1}{2} \sum_{\mathbf{i},\mathbf{j}} V_{\mathbf{ij}} (\langle n_{\mathbf{i}} \rangle \langle n_{\mathbf{j}} \rangle - \left| \langle c_{\mathbf{j}}^{\dagger} c_{\mathbf{i}} \rangle \right|^2) \qquad (3)$$

which consists of a random potentiel (Hartree terms) $w_{\mathbf{i}}^{eff} = w_{\mathbf{i}} + \sum_{\mathbf{j} \neq \mathbf{i}} V_{\mathbf{ij}} \langle n_{\mathbf{j}} \rangle$ and random hopping terms (Fock terms) $t_{\mathbf{ij}}^{eff} = t_{\mathbf{ij}} + V_{\mathbf{ij}} \langle c_{\mathbf{j}}^{\dagger} c_{\mathbf{i}} \rangle$. The quantities $\langle n_{\mathbf{j}} \rangle$ and $\langle c_{\mathbf{j}}^{\dagger} c_{\mathbf{i}} \rangle$ are calculated self consistently.

It is important to note that the HF current can be calculated alternatively by using $I(\Phi) = -\frac{1}{2\pi} \frac{\partial E_{hf}(\Phi)}{\partial \Phi}$ or $I(\Phi) = \langle J_x \rangle$. Although $E_{hf}$ is the expectation value $\langle \mathcal{H} \rangle$ in the HF groundstate which contains the flux-dependance both in $\mathcal{H}$ and in the HF wavefunction, the self-consistency implies that $\delta \langle \Psi | \mathcal{H} | \Psi \rangle$, where the variation is made on the wavefunction *only*, is identically vanishing. A comparison between these two quantities provides then a useful check of the convergence of the self-consistent procedure.

HF and exact results both obtained on the same $4 \times 4$ cluster are compared in fig.1. We observe a rather good agreement between them and the same trends when V is applied. This provides then some evidence that the HF approach gives qualitatively correct results.

We have observed for the $4 \times 4$ system a tendency of the localization length to decrease with increasing repulsion in agreement with the fact that the current itself decreases. So far, our data suggest that the interaction do not enhance the current in the *ballistic* regime. This motivates the study of bigger systems where the diffusive regime can be more easily realized.

¿From now on we shall consider larger systems of size $6 \times 6$ and $8 \times 8$ which can be handled only by using the HF approximation. It is important to increase simultaneously $L_y$ with $L_x$ since, at fixed density, the number of energy level crossings increases with the number of channels. We show in fig.2 the current versus $\phi$ for the clean system to illustrate this point. Level crossings lead to discontinuities of the current at some particular values of the flux. The sign of the current depends on the parity of the number of electrons[16]. We now compare the effect of short range and Coulomb repulsions in the presence of disorder. In fig.3 and fig.4 we have plotted $I_1$ and the width of the distribution $\delta I = \sqrt{I_2 - I_1^2}$ versus $\phi$. The calculations are done for $W = 3$, and $V = 0$ and 0.5. Note that the sign of the current has not changed in comparison with the clean case. First, we observe that the width of the distribution decrease with the interaction, the effect being larger in the case of short range repulsion. Another important point is the large effect on the first moment. We clearly observe that, in this diffusive regime, the Coulomb interaction enhances the current, while the screened interaction has a weaker effect. Note that the effect of the Coulomb repulsion is still small in the sense that it is far from cancelling out completely the drastic effect of the disorder potential to restore the clean current. However, we have to keep in mind that the rings considered here contain a small number of channels and, as shown in fig.2, the number of level crossings is still quite small. Considering this last point we argue that the effect observed here is in fact quite significant. We expect the effect to become much larger when the number of channels (and hence of level crossings) becomes larger. Note that similar increase of the currents have also been obtained recently by exact diagonalizations of $4 \times 4$ clusters in the *large W regime*[9].

In conclusion, the HF results are in qualitatively good agreement with the exact calculations. We find that, in the case of spinless fermions, the Coulomb interaction enhances the current, while a screened interaction has a weaker effect. The increase (decrease) of the current is always consistently related to the increase (decrease) of the localization length i.e. to a delocalization (localization) of the electronic wavefunctions in average. Qualitatively we expect a fraction of the Fermi gas to become localized at the minima of the potential and, because of the interaction, to screen out these minima for the rest of the electrons. This physical picture is particularly clear in the HF equations themselves where a new effective potential is generated by the interaction. Whether sufficient screening can be self-consistently generated depends on the nature of the interaction between fermions. Our work strongly suggests that a long range interaction is more effective to smooth out the impurity potential and hence to increase the current. We also argue that this effect will become stronger and stronger as the number of channels is increased. A detailed study of the role of the transverse dimension is left for future work.

We gratefully acknowledge discussions with G. Montambaux, T. Giamarchi, D. Shepelianski and C. Sire. We also thank the *Institut du Developpement et des Ressources en Informatique Scientifique (IDRIS)*, Orsay (France) for allocation of CPU time on the Crays C98 and C94. D.P. acknowledges support from the EEC Human Capital and Mobility Program under grant CHRX-CT93-0332. *Laboratoire de Physique Quantique (Toulouse)* is *Unité Associée No. URA505 du CNRS*.

* Electronic address: georges@irsamc2.ups-tlse.fr
** Electronic address: didier@irsamc2.ups-tlse.fr

FIG. 1. First moment $I_1$ (full symbols) and second moment $I_2$ (open symbols) of the current distribution at $\phi = 1/4$ for W=1 (a) and W=3 (b). Squares (triangles) correspond to the HF (exact) current $\langle J_x \rangle$. The average is taken over $N_r = 500$ realisations of the disorder in the ED case and over $N_r = 1000$ in the HF approximation. Error bars of order $\delta I/N_r$ are included unless smaller than the size of the dots.

FIG. 2. First moment $I_1$ versus $\phi$ at $V=0$ and $W=0$ for 4×4, 6×6 and 8×8 systems.

FIG. 3. Moments of the current of a 6×6 site system for $W = 3$. Each symbol represents an average over 500 configurations of the disorder. (a) First moment $I_1$ for $V = 0$ (triangles), a short range interaction $V = 0.5$ (squares) and a Coulomb repulsion $V = 0.5$ (pentagons). (b) width of the distribution $\delta I$ for $V = 0$ (triangles), a short range interaction $V = 0.5$ (squares) and a Coulomb interaction $V = 0.5$ (pentagons). Typically, the error bars corresponding to the $I_1$ data (not shown) are of the order of $\frac{1}{25}I_2$.

FIG. 4. Same as fig.3 for a 8×8 site system.

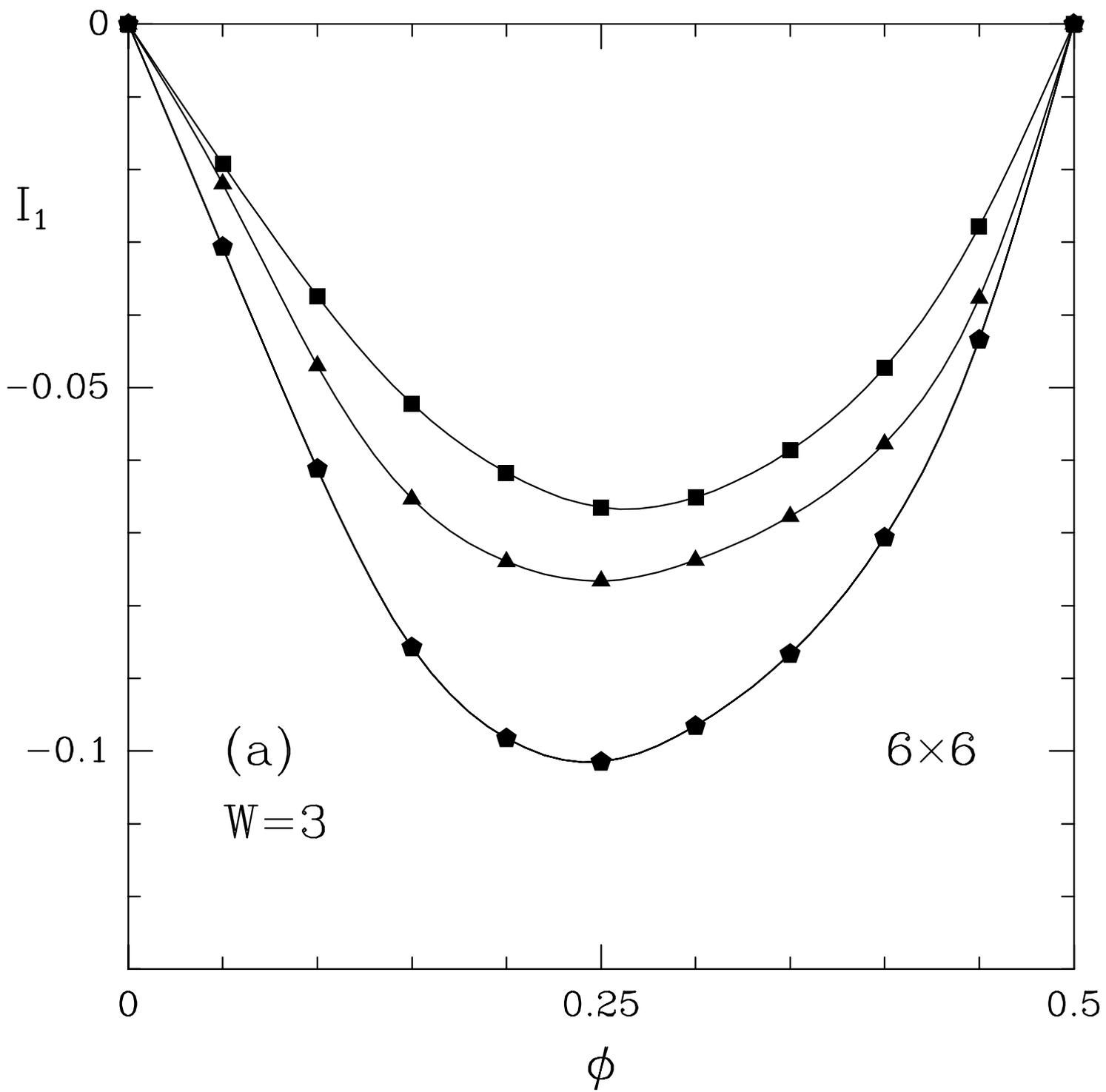

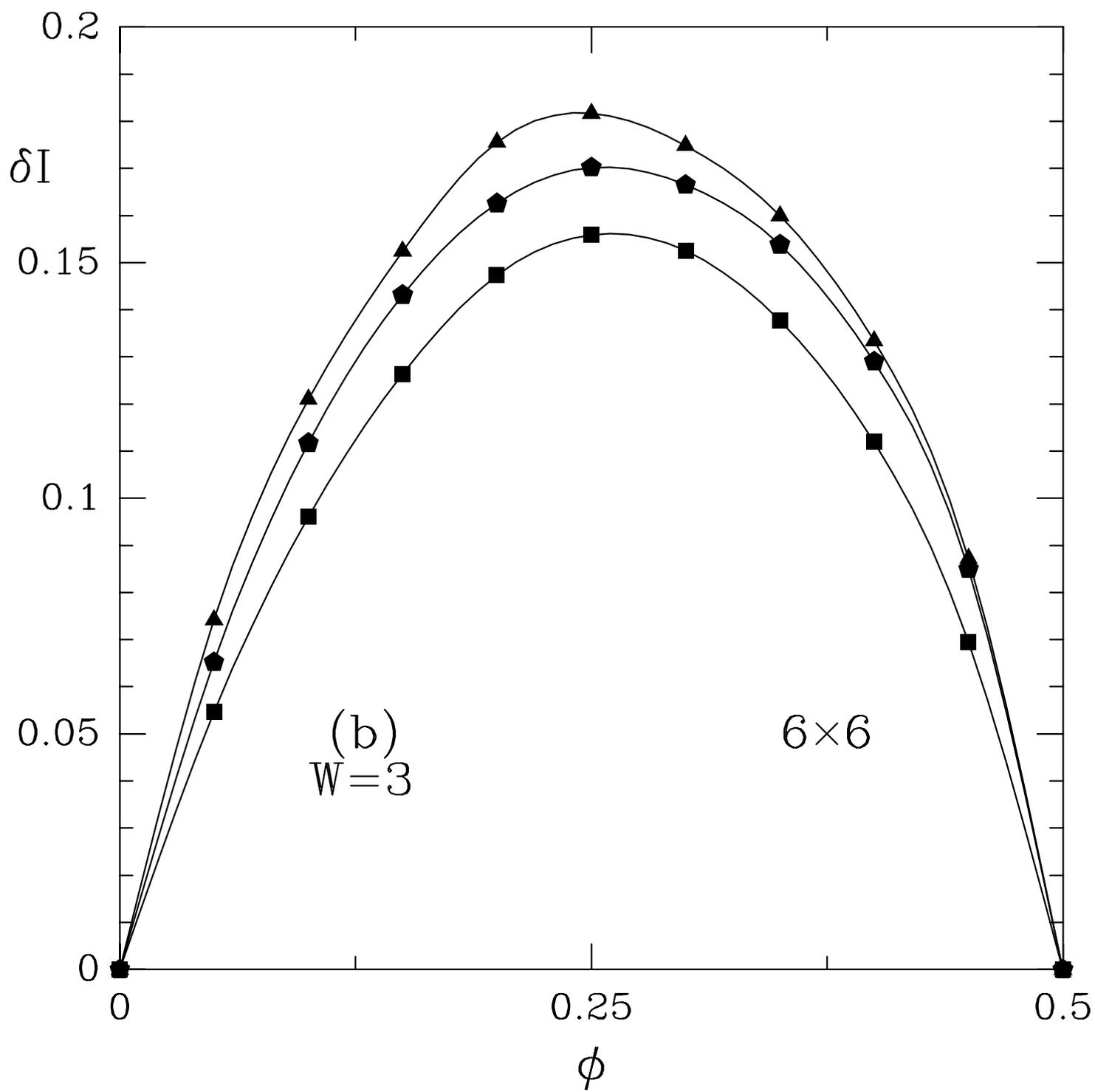

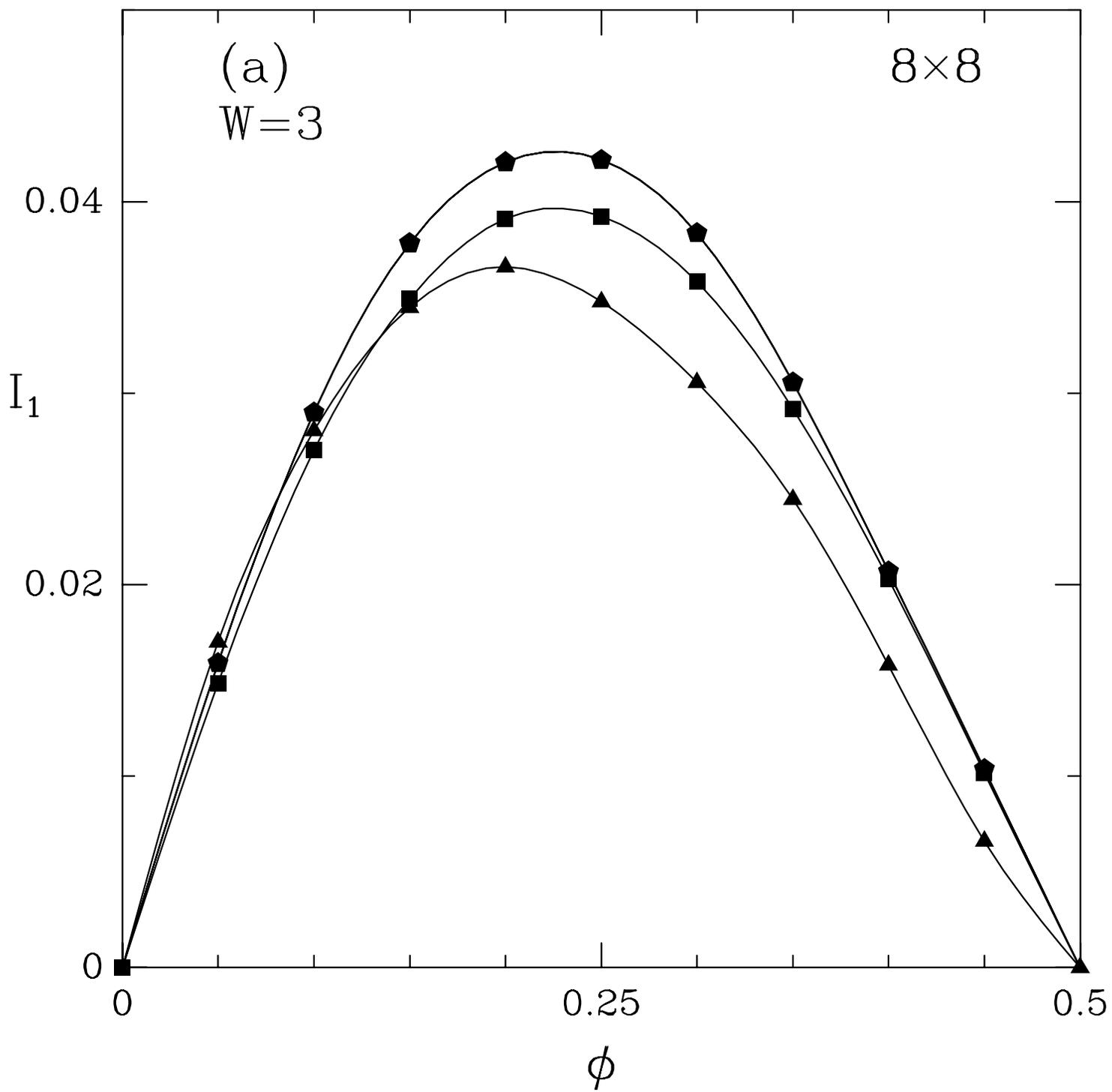

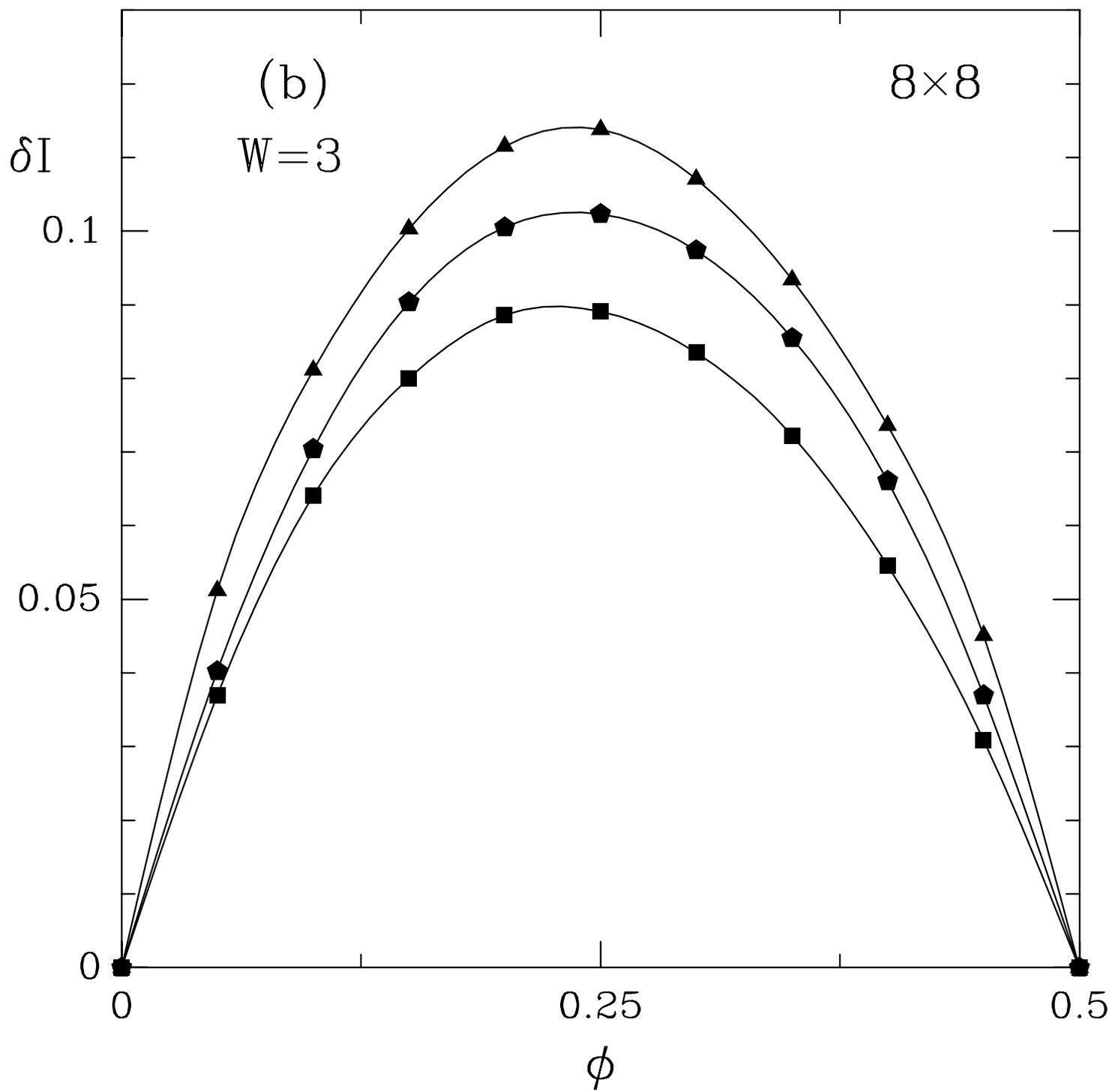